\documentclass[twocolumn,english,aps,prl]{revtex4}
\usepackage[T1]{fontenc}
\usepackage[latin9]{inputenc}
\usepackage{color}
\usepackage{amsmath}
\usepackage{amssymb}
\usepackage{graphicx}
\usepackage{natbib}
\usepackage{braket}
\usepackage{psfrag}
\usepackage{tikz}
\usepackage{physics}
\usepackage{cancel}

\usepackage[colorlinks=true,urlcolor=blue,citecolor=blue,linkcolor=blue]{hyperref}

\usepackage{babel}

\usepackage[normalem]{ulem}

\newcommand{\Heff}{\hat{H}_{\rm eff}}

\newcommand{\ophi}{\omega_\Phi}
\newcommand{\phiphi}{\varphi_\Phi}

\begin{document}

\title{Gate-efficient simulation of molecular eigenstates on a quantum computer}

\author{M. Ganzhorn, D.J. Egger, P. Barkoutsos, P. Ollitrault, G. Salis, N. Moll, A. Fuhrer, P. Mueller, S. Woerner, I. Tavernelli and S. Filipp}
\affiliation{IBM Research Zurich, S\"aumerstrasse 4, 8803 R\"uschlikon, Switzerland}

\begin{abstract}
A key requirement to perform simulations of large quantum systems on near-term quantum hardware is the design of quantum algorithms with short circuit depth that finish within the available coherence time. A way to stay within the limits of coherence is to reduce the number of gates by implementing a gate set that matches the requirements of the specific algorithm of interest directly in hardware. Here, we show that exchange-type gates are a promising choice for simulating molecular eigenstates on near-term quantum devices since these gates preserve the number of excitations in the system. Complementing the theoretical work by Barkoutsos et al. [PRA 98,
022322 (2018)], we report on the experimental implementation of a variational algorithm on a superconducting qubit platform to compute the eigenstate energies of molecular hydrogen. We utilize a parametrically driven tunable coupler to realize exchange-type gates that are configurable in amplitude and phase on two fixed-frequency superconducting qubits.  With gate fidelities around 95\% we are able to compute the eigenstates within an accuracy of  50 mHartree on average, a limit set by the coherence time of the tunable coupler. %Future improvements in coupler coherence to at least 1 $\mu$s along with more qubits will allow for the computation of larger molecules, such as water or even molecular nitrogen \textcolor{blue}{(@Niko/Ivano: please comment on the nitrogen)}.%In a proof of principle experiment, to calculate the two lowest-lying eigentstate of molecular hydrogen.
\end{abstract}

\date{\today}

\maketitle

%\section{Introduction}
The simulation of the electronic structure of molecular and condensed matter systems is a challenging computational task as the cost of resources increases exponentially with the number of electrons when accurate solutions are required. 
%size of the system. Efficiently simulating the problem is one of the main concerns of chemistry, condensed matter physics and material science. 
With the tremendous improvements in our ability to control complex quantum systems this bottleneck may be overcome by the use of quantum computing hardware~\cite{Feynman1982}. In theory, various algorithms for quantum simulation have been designed to that end, including quantum phase estimation~\cite{Kitaev1995} or adiabatic algorithms~\cite{Babbush2014}. With these algorithms the challenges for practical applications lie in the efficient mapping of the electronic Hamiltonian onto the quantum computer and in the required number of quantum gates that remains prohibitive on current and near-term quantum hardware~\cite{Wecker2014} without quantum error correction schemes~\cite{Fowler2012}. On the other hand, variational quantum eigensolver (VQE) methods~\cite{Peruzzo2014,Moll2018} can produce accurate results with a small number of gates~\cite{Barkoutsos2018} using for instance algorithms with low circuit depth~\cite{Kivlichan2018} and do not require a direct mapping of the electronic Hamiltonian onto the hardware. Moreover, such algorithms are inherently robust against certain errors~\cite{Barkoutsos2018,McClean2016a,Babbush2018} and are therefore considered as ideal candidates for first practical implementations on non error-corrected, near-term quantum hardware. 

%Indeed, 
Recently, the molecular ground state energy of hydrogen and helium have been computed via VQE in proof of concept experiments using NMR quantum simulators~\cite{Lanyon2010,Du2010,Li2011}, photonic architectures~\cite{Peruzzo2014} or nitrogen-vacancy centers in diamond~\cite{Wang2015}. Although very accurate energy estimates are obtained, quantum simulation of larger systems remains an intractable problem on these platforms because of the difficulties arising in scaling them up to more than a few qubits. For this reason trapped ions~\cite{Monroe2013,Zhang2017,Bernien2017,Hempel2018} and superconducting qubits~\cite{Neill2017,Otterbach2017,QuantumExperience} have become promising candidates to carry out VQE-based quantum simulations in particular for quantum chemistry applications. For instance, the ground state energies of molecules like $\rm H_2$~\cite{omalley_scalable_2016,Kandala2017,Colless2018}, $\rm LiH$ and $\rm BeH_2$~\cite{Kandala2017}, as well as the energy spectrum of the four eigenstates of $\rm H_2$~\cite{Colless2018}, have already been measured on general purpose superconducting qubit platforms. In these experiments, a heuristic approach based on gates already available in hardware, such as C-Phase, CNOT or bSWAP, is employed. However, computing larger molecules with more orbitals in the active computational space becomes impractical with this method. Without further constraints, the dimension of the Hilbert space accessed via the parameterized gate sequences in the VQE grows exponentially with the number of required qubits $N$. The probability to reach the wanted ground state decreases accordingly. It is, thus, important to use a set of entangling gates that matches the specifics of the problem~\cite{Barkoutsos2018}. For quantum chemistry calculation, each qubit typically represents the population of an electronic orbital~\cite{Jordan1928,Bravyi2002}. Since the number of electrons $n_e$ for a given molecular system or a chemical reaction is constant, the total number of qubit excitations is also constant. 
Exchange-type interactions, which preserve the number of excitations on the qubit processor are, therefore, better suited than other two-qubit gates to compute molecular eigenstates~\cite{Barkoutsos2018,Motzoi2017}. In fact, using only excitation-preserving gates constrains the accessible state space to a subspace of the full $2^N$-dimensional Hilbert space: only the $\binom{N}{n_e}$-dimensional manifold with $n_e$ electrons is explored in VQE. Such a reduced search space is beneficial for both the construction of a reduced molecular Hamiltonian~\cite{Moll2018} as well as for the expansion of the trial wavefunction~\cite{Barkoutsos2018}.

In this paper, we show an efficient and scalable approach to compute the energy spectrum of molecules using exchange-type two-qubit gates. We demonstrate in simulation that the circuit depth required to achieve chemical accuracy in a VQE algorithms can be significantly reduced by using exchange-type gates, which would allow the simulation of larger quantum systems on near-term quantum hardware. We implement such an exchange-type gate based VQE algorithm on a hardware platform consisting of two fixed-frequency superconducting qubits coupled via a tunable coupler~\cite{McKay2016,Roth2017} and determine the ground state energy of molecular hydrogen. Finally, we derive the excited states of molecular hydrogen from the measured ground state using the equation-of-motion (EOM) approach~\cite{Rowe1968}, which complements the quantum subspace expansion in~\cite{McClean2016,Colless2018}. The EOM approach is a well known and accurate quantum chemistry method, although not widely used since computing the matrix elements of its characteristic pseudoeigenvalue system of equations is an exponentially hard computational problem~\cite{Dreuw2005}. On a quantum processor we can efficiently measure these matrix elements and classically derive the excited state energies. \\
%We efficiently compute the excited state of molecular hydrogen using a classical equation-of-motion approach, a well known method in quantum chemistry with known ..... \textcolor{blue}{(@Pauline,Ivano: mention some benefits...)}~\cite{Rowe1968}. It derives the excited state energies from the ground state energy obtained by VQE, completing the quantum subspace expansion in~\cite{McClean2016,Colless2018}. \\

\noindent
\textbf{Results}\\

\textbf{Efficient circuit design with exchange-type gates.} Efficient VQE algorithms for the solution of electronic structure problems in quantum chemistry are formulated in a second quantization framework~\cite{Babbush2018,Moll2018}. % \textcolor{blue}{The Hamiltonian is represented as a sum of one-body and two-body terms, which are expressed as a linear combination of electron creation and annihilation operators. (Ivano/Panos please confirm)}
In this context, the molecular Hamiltonian is represented as a sum of one and two-body terms and then mapped to the qubit space using a fermion-to-qubit transformation, like the Jordan-Wigner~\cite{Jordan1928} or the parity mapping transformation~\cite{Bravyi2017}.
%, which are expressed as a linear combination of electron creation and annihilation operators. (Ivano/Panos please confirm)}
Suitable trial states for VQE can be computed with a unitary coupled cluster (UCC) ansatz~\cite{Taube2006}. However, the complexity of the trial state generation increases rapidly with the size of the system, precluding the simulation of larger systems on near-term quantum hardware using the UCC ansatz~\cite{Moll2018,Barkoutsos2018}. %This deficiency can be mitigated with a first-order Trotter decomposition of the clustering operators and a reduction of the Hilbert space size using a suitable active space~\cite{Panos2018}. 
Alternatively, a heuristic generation of trial states by a sequence of unitary operations can also be efficiently implemented on near-term quantum hardware~\cite{Barkoutsos2018}. In the original formulation~\cite{Kandala2017}, the heuristic trial wavefunction was generated in the full Fock-space, thus including states with all possible numbers of electrons.  With each qubit being mapped to the population of an electronic orbital this corresponds to a Hilbert space spanned by the $2^N$ basis states $\{i_1,i_2,...,i_N\}$ with $i_k = 0,1$. However, in quantum chemistry the solution of interest lies usually in the sector of the Hilbert space with a well defined number of electrons $n_e$, i.e. a space spanned by all basis states with $\sum_k i_k =n_e$. It is therefore advantageous to generate trial states that are part of this restricted subspace by designing circuits that conserve the total number of excitations over the entire qubit register. 

The simplest method to do this is to prepare the initial state by exciting $n_e$ qubits e.g. $|1_1,1_2,...,1_{n_e},0,...,0\rangle$ and apply only gates that exchange excitations between qubits without creating ($\hat{\sigma}^+$) or annihilating ($\hat{\sigma}^-$) new excitations. The corresponding two-qubit operation is an exchange-type gate generated by $(\hat{\sigma}^+ \hat{\sigma}^- + h.c.)$. The size of the restricted subspace is then given by $\binom{N}{n_e} \leq 2^N$. Close to half-filling with $n_e \approx N/2$, the advantage is small since $\binom{N}{n_e} \approx 2^{N/2}$. For many molecules however, the number of electrons is typically $n_e \approx N/10$~\cite{chem_database} and the size of the restricted subspace $\binom{N}{n_e} \approx (N/n_e)^{n_e}$ is significantly smaller than that of the full Hilbert space.% with $2^N$ states.

\begin{figure}[t]
\includegraphics[width = 0.48\textwidth]{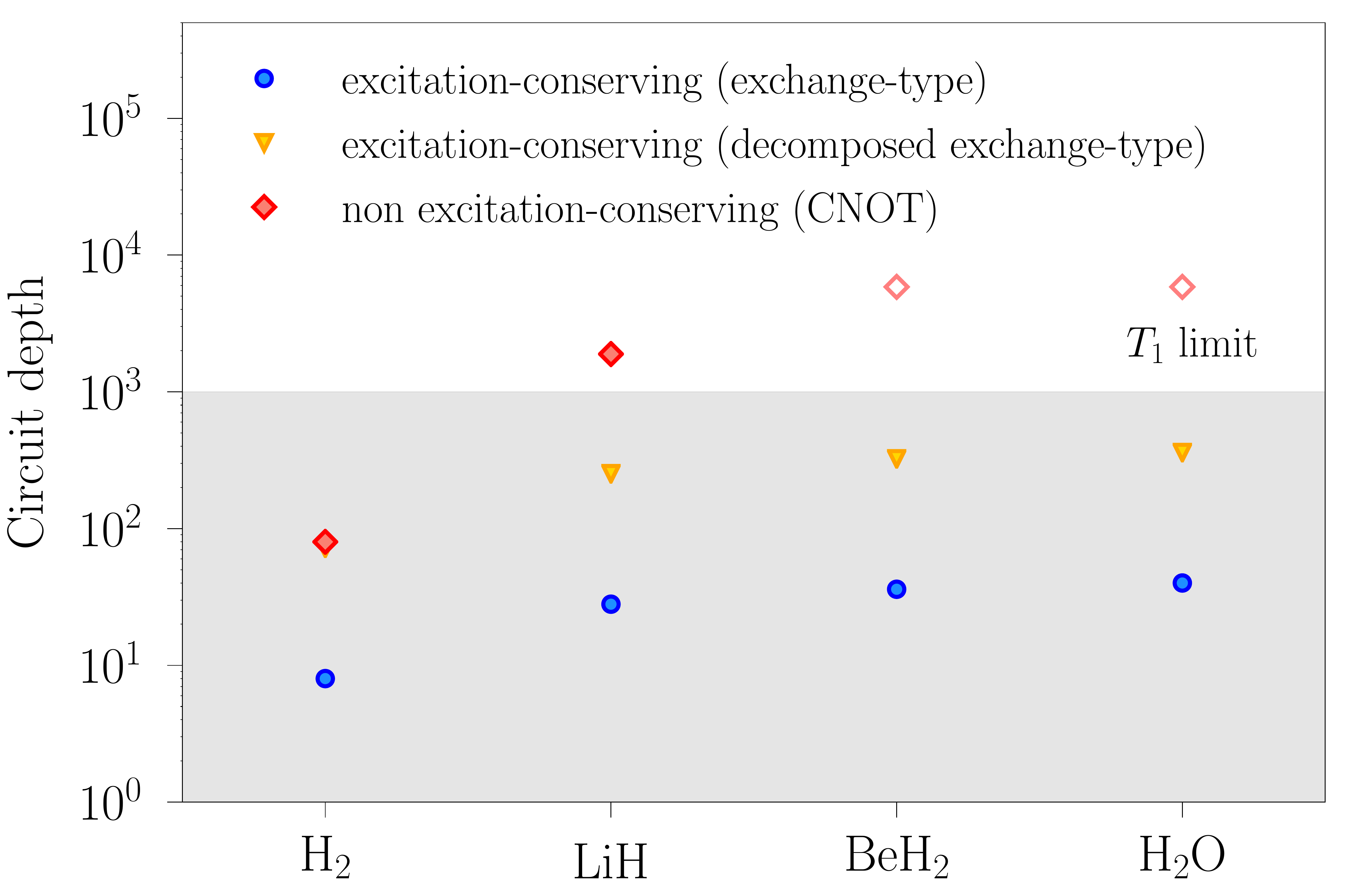}
\caption{Circuit depth required to achieve chemical accuracy for the ground state energy with a VQE algorithm for the $\rm H_2$, $\rm LiH$, $\rm BeH_2$ and $\rm H_2O$ molecules. Non excitation-conserving circuits based on CNOT gates (red squares) are compared to excitation-conserving circuits based on exchange-type gates (blue circles) and a decomposition thereof into CNOT's (yellow triangles). In some cases, only a lower boundary to the circuit depth could be estimated (empty symbols). Bounded by the $T_1$ time in the currently available hardware, only circuits within the grey region can be practically implemented without error mitigation or reduction schemes (see text). \label{fig:theory}}
\end{figure}

In a VQE simulation, the size of the explored subspace is directly connected to the circuit depth required to reach a certain accuracy. Assuming error free gates and using the minimal basis set of atomic orbitals typically used in quantum chemistry~\cite{Pople1970}, we estimate the circuit depth required to achieve chemical accuracy in a VQE simulation of the molecules $\rm H_2$, $\rm LiH$, $\rm BeH_2$ and $\rm H_2O$ (see Fig.\ref{fig:theory} and Supplementary material). Heuristic non excitation-conserving circuits, based e.g. on CNOT gates~\cite{Kandala2017}, can in principle achieve chemical accuracy for these molecules. However, the required circuit depth becomes prohibitively large for molecules bigger than $\rm H_2$ as the circuit runtime exceeds the best relaxation times $T_1 \sim 100$ $\mu$s currently available in superconducting hardware. %Consequently, VQE simulations using non excitation-conserving circuits seem currently impossible to realize without further reduction schemes~\cite{Kandala2017} or error mitigation techniques~\cite{Temme2016,Kandala2018}. 
On the other hand, circuits based on excitation-conserving exchange-type gates require a much shorter circuit depth and achieve chemical accuracy for all studied cases within the $T_1$ limit without further amendments (Fig.\ref{fig:theory}). Clearly, the wanted excitation-preserving two-qubit gate could be decomposed into the available universal gate set~\cite{Chuang2000}, e.g. using CNOT gates. But this comes at the expense of a ninefold increase in circuit depth (Fig.\ref{fig:theory}) that can be avoided by using application specific hardware and gates. We note that additional reduction schemes can be used to minimize the number of qubits as demonstrated in Ref.~\cite{Kandala2017} for $\rm H_2$, $\rm LiH$, $\rm BeH_2$ and as discussed in the following for the proof-of-principle determination of the eigenspectrum of $\rm H_2$.
%In the following, we discuss the experimental implementation of a exchange based VQE algorithm using a tunable coupler architecture~\cite{McKay2016,Roth2017}. 

\begin{figure}[hbt!]
\includegraphics[width = 0.48\textwidth]{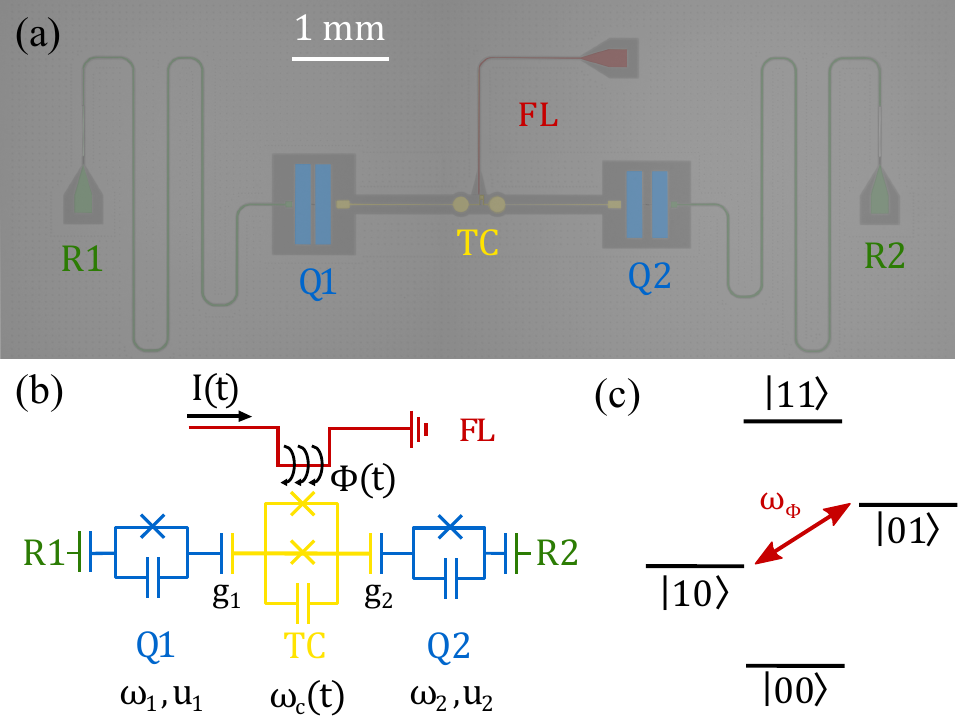}
\caption{(a) Optical micrograph and (b) circuit scheme of the device consisting of two fixed-frequency transmons (Q1, Q2) capacitatively coupled to a flux-tunable transmon acting as tunable coupler (TC). The tunable coupler is controlled by a flux line (FL) providing a current $I(t)$ and a consequent flux $\Phi(t) = \Phi_{\text DC} + \delta \cos(\ophi t + \phiphi)$ threading the SQUID-loop of the coupler. Each of the fixed-frequency qubits is coupled to an individual readout resonator (R1, R2). (c) Level diagram of the device. Here, $\ket{n_1 n_2}$ denotes the state of the combined system with the qubit excitation number $n_{1,2}$. Modulation of the magnetic flux $\Phi(t)$ at the qubits difference frequency $\omega_{\Phi} = \omega_1-\omega_2$ drives the transition between $\ket{10}$ and $\ket{01}$.} %The states $\ket{100}$ and $\ket{010}$ are separated by the difference frequency $\sim \omega_1 - \omega_2$ whereas the states $\ket{000}$ and $\ket{110}$ are separated by the sum frequency $\sim \omega_1+\omega_2$. Additionally, the most important sideband transitions between qubits and coupler are shown. }
\label{fig:setup}
\end{figure}

%\section{Experiment - Tunable coupling architecture}
%\label{sec:exp}

%\subsection{Creating a exchange primitive $\hat U_\text{exchange}$}
%\label{subsec:chem_gate}

\noindent
\textbf{Hardware implementation of exchange-type gate.} An exchange-type gate primitive can naturally be realized in a tunable coupler architecture (Fig.~\ref{fig:setup})~\cite{McKay2016,Roth2017}. The device consists of two fixed-frequency transmon qubits Q1 and Q2 linked via a tunable coupler (TC), i.e. a frequency-tunable transmon. Spectroscopic measurements of the device yield qubit frequencies $\omega_{1,2}$, capacative coupling strengths $g_{1,2}$ and decay rates as summarized in Table~\ref{tab:params}. An exchange-type coupling between the computational qubits Q1 and Q2 is achieved by parametric modulation of the TC frequency $\omega_\text{c}(t) = \omega_\text{c}^0\sqrt{\abs{\cos(\pi\Phi(t)/\Phi_0)}}$ \cite{McKay2016, Roth2017}. Threading a magnetic flux $\Phi(t) = \Phi_{\text DC} + \delta \cos(\ophi t + \phiphi)$ with $\ophi = \omega_1-\omega_2$ through the SQUID loop of the TC implements the effective Hamiltonian \cite{Roth2017}
\begin{eqnarray}
\Heff = \Omega_{\rm eff} \left(XX+YY\right),
\label{eq:heff_iswap}
\end{eqnarray}
with the set of Pauli operators $\{X,Y,Z\} \equiv \{\hat{\sigma}_x,\hat{\sigma}_y, \hat{\sigma}_z\}$. It describes an exchange-type interaction between $\ket{10}$ and $\ket{01}$ at a rate $\Omega_{\rm eff} (\Phi_{\text DC},\delta)$ (Fig. \ref{fig:setup}(c)). The resulting two-qubit gate operation is described by the unitary operator 
\begin{align}
\hat U_\text{EX} (\theta,\varphi) =\begin{pmatrix} 
1 & 0 & 0 & 0 \\
0 & \cos\theta/2 & ie^{i\varphi}\sin\theta/2 & 0 \\ 
0 & ie^{-i\varphi}\sin\theta/2 & \cos\theta/2 & 0 \\ 
0 & 0 & 0 & 1 \\ 
\end{pmatrix}
\end{align}
Here, $\theta = \pi \tau/\tau_{\pi}$ is controlled by the length $\tau$ of the tunable coupler drive pulse and $\tau_{\pi} = 170$ ns is the length of an iSWAP gate, which completely transfers an excitation from one qubit to the other. The phase $\varphi = \varphi_{\Phi} + \varphi_{\rm off}$ is controlled by the phase $\varphi_{\Phi}$ of the tunable coupler drive with $\varphi_{\rm off}$ being a global phase offset determined by the actual relative phases of the microwave drives.

\begin{table}
\centering
\resizebox{0.47\textwidth}{!}{%
\begin{tabular}{lcccccc}
\hline\hline
   & $\omega/2\pi$ (GHz) & $\alpha/2\pi$ (MHz)    		& $g/2\pi$ (MHz) 				& $T_1$ ($\mu$s) & $T_2$ ($\mu$s) & $T_2^*$ ($\mu$s) \\
Q1 & $4.959 $     			 & $-236 \pm 1$				  & $80 \pm 2$					  & $78 \pm 1$     & $97 \pm 1$     & $86 \pm 1$			 \\
Q2 & $6.032 $     			 & $-308 \pm 1$  				& $120 \pm 2$					  & $23 \pm 1$     & $23 \pm 2$     & $13 \pm 1$ 			 \\
TC & $7.370 $     			 & $-255 \pm 6$ 				& n/a        						& $6.3 \pm 0.7$     & $0.08 \pm 0.01$ 				& $0.02 \pm 0.01$ \\
\hline\hline
\end{tabular}}
\caption{Device parameters of two fixed-frequency transmons (Q1, Q2) coupled via a flux-tunable transmon (TC) as shown in Fig.~\ref{fig:setup}. The qubits exhibit frequencies $\omega/2\pi$, anharmonicities $\alpha/2\pi$, and capacitive couplings $g/2\pi$ between qubits and TC at zero flux bias ($\Phi_{\text DC}=0$). The relaxation time $T_1$, spin-echo coherence time $T_2$ and Ramsey coherence time $T_2^*$ are measured at the flux bias point $\Phi_{\text DC}= 0.195\, \Phi_0$ (see Supplementary material for details and additional measurements). %$E_{\rm J}/E_{\rm c}$ is infered from the frequencies and anharmonicities as listed above. 
\label{tab:params}}
\end{table}

\begin{center}
\begin{figure}[hbt!]
\includegraphics[width=0.48\textwidth]{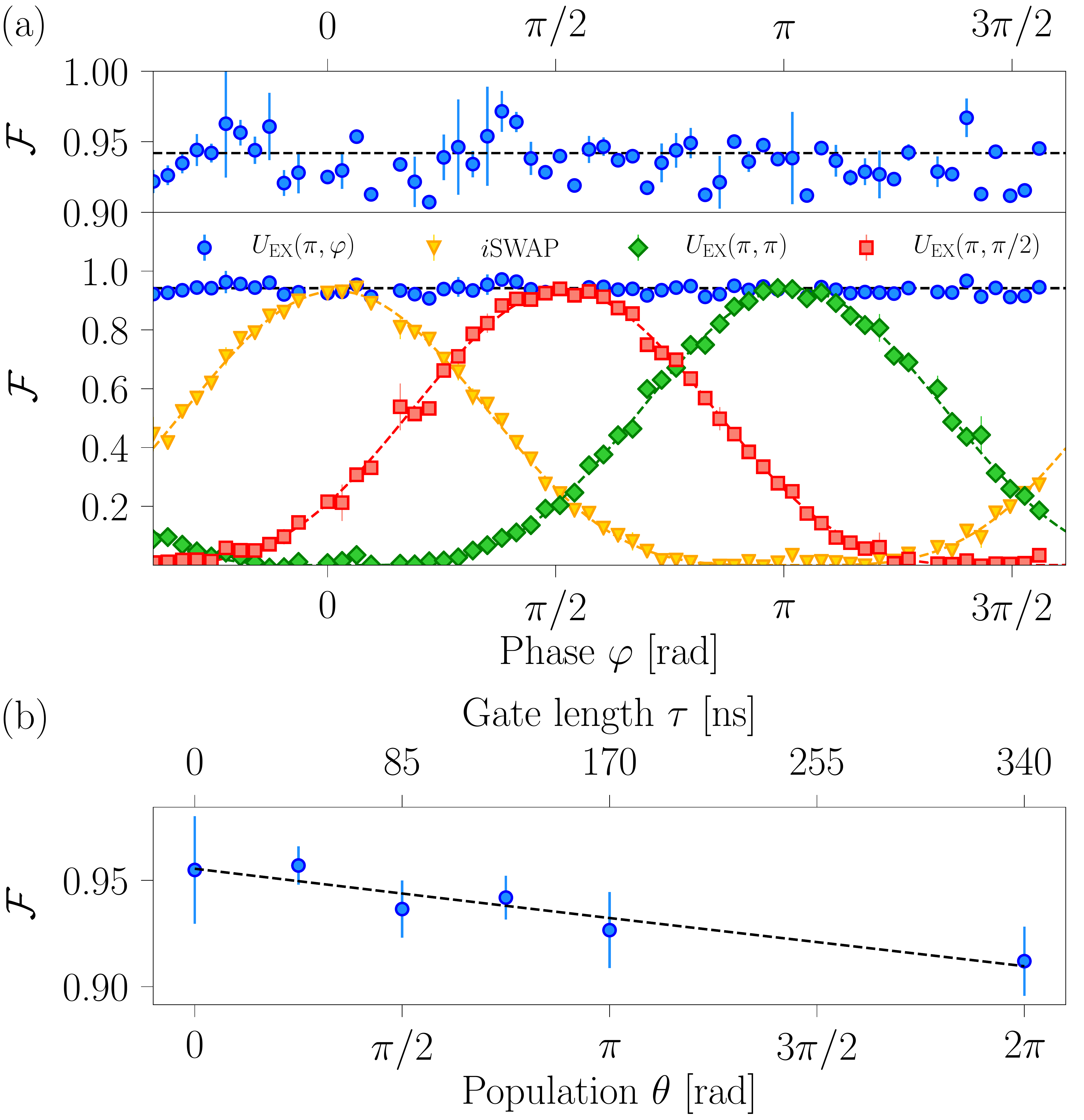}
\caption{Quantum process tomography of the chemistry gate $\hat{U}_\text{EX}(\theta,\varphi)$. (a) Gate fidelities $\mathcal{F}$ as a function of $\varphi$ for $\theta = \pi$. The bottom panel shows the gate fidelities calculated from the overlap of the measured process matrices $\chi_{\rm meas} (\varphi)$ with the ideal process matrix $\chi_{\rm ideal}$ of a $\hat{U}_\text{EX}(\pi,\varphi)$ (blue dots), iSWAP (orange triangles), $\hat{U}_\text{EX}(\pi,\pi/2)$ (red squares) and $\hat{U}_\text{EX}(\pi,\pi)$ (green diamonds) gate operation. The top panel shows the gate fidelities with respect to $\hat{U}_\text{EX}(\pi,\varphi)$. Black dashed lines depicts the average gate fidelity for $\hat{U}_\text{EX}(\pi,\varphi)$ (see text). Colored dashed lines are a fit to equation \ref{eq:fid}. (b) Gate fidelities $\mathcal{F}$ as a function of $\theta$ where the phase $\varphi_{\rm opt}$ is tuned to maximize QPT fidelity. Dashed line is a fit with an exponential decay function with a decay time of $6.7$ $\mu$s.
\label{fig:control}}
\end{figure}
\end{center}

To benchmark the efficiency of the exchange-type gate primitive, we perform quantum process tomography (QPT) of $\hat{U}_\text{EX}$ as function of $\varphi$ for a fixed $\theta=\pi$. The overlap of the measured process matrix $\chi_{\rm meas}(\varphi)$ with an ideal process matrix $\chi_{\rm ideal}$ yields the gate fidelity $\mathcal{F}=\text{Tr}(\chi_{\rm meas}(\varphi)\chi_{\rm ideal})$. If the measured process matrices are compared with the ideal process matrix of a $\hat{U}_\text{EX}(\pi,\varphi)$ operation, the gate fidelity is constant over $\varphi$ with an average of $\mathcal{F} = 94.2\pm1.5\%$ [Fig. \ref{fig:control}(a)]. However, if the measured process matrices are compared with the ideal process matrix of $\hat{U}_\text{EX}(\pi,0)$, equivalent to an iSWAP gate operation, the gate fidelity is phase dependant. A fit with the analytic expression 
\begin{equation}
\mathcal{F}_{\rm ana}=\mathcal{F}_0 | e^{-2i(\varphi-\varphi_0)}(1+e^{i(\varphi-\varphi_0)})^4| 
\label{eq:fid}
\end{equation}
yields a maximum  gate fidelity of $\mathcal{F}_0 = 93.2 \pm 0.5 \%$ achieved for $\varphi_0 = 3 \pm 5$ mrad (Fig.~\ref{fig:control}(a)). Similarly, a comparison with the ideal process matrix of $\hat{U}_\text{EX}(\pi,\pi/2)$ and $\hat{U}_\text{EX}(\pi,\pi)$ yields a maximum gate fidelity at $\varphi_0 = 1.574 \pm 0.007$ rad and $\varphi_0 = 3.155 \pm 0.006$ rad, respectively. It should be noted that the gate fidelity estimation via QPT is subject to state preparation and measurement (SPAM) errors. Other techniques like randomized benchmarking are robust against such SPAM errors, but are mostly limited to gates from the Clifford group. For an iSWAP as a two-qubit gate primitive, we find an error per gate of $2.5 \%$ via randomized benchmarking (see Supplementary material).%Recently randomized benchmarking has been extended to some two qubit non-Clifford gates~\cite{Cross2016}, but gate fidelity estimations for non-Clifford gates like $\hat{U}_\text{EX}(\theta\neq\pi, \varphi)$ are to our knowledge still pending. 

Furthermore, we perform QPT of $\hat{U}_\text{EX}$ as function of $\theta$, i.e. for different lengths $\tau$ of the drive pulse on the tunable coupler. Comparing the measured process matrices with the ideal process matrix of $\hat{U}_\text{EX} (\theta,\varphi_\text{opt})$ yields gate fidelities ranging from $\mathcal{F} = 96\pm2.5\%$ (for small $\theta$) to $\mathcal{F} = 91\pm1.5\%$ (for large $\theta$) (Fig. \ref{fig:control}(b)). Here, the phase $\varphi_{\rm opt}$ is calibrated to maximize fidelity. The observed decrease of gate fidelity with increasing $\theta$, i.e. longer pulse length $\tau$, can be fitted to an exponential function with a decay time of $6.7$ $\mu$s, close to the measured relaxation time $T_1 = 6.3$ $\mu$s of the TC (see Table \ref{tab:params}). 

%\subsection{Computing molecular spectra with SWAP primitive: The hydrogen molecule}

%In the following, the ground state energy of molecular hydrogen is computed with a VQE algorithm using the SWAP primitive described above. Based on the calculation of the ground ground state, we compute in a second step the excited state of molecular hydrogen using an equation of motion (EOM) approach.

\noindent
\textbf{Computing molecular spectra with exchange-type gates.} To demonstrate the usefulness of this gate, we now compute the ground state and the three excited states of molecular hydrogen. Using a parity mapping transformation~\cite{Bravyi2017}, we map the fermionic second-quantized Hamiltonian of molecular hydrogen to the two-qubit Hamiltonian
\begin{eqnarray}
\hat{H}_{\rm H_2} = \alpha_0 II + \alpha_1 ZI + \alpha_2 IZ + \alpha_3 ZZ + \alpha_4 XX
\label{eq:Hamiltonian}
\end{eqnarray}
where $\alpha_i$ denote pre-factors that are classically computed as a function of the bond length of the molecule in the STO-3G basis (see Supplementary material). 

\begin{center}
\begin{figure}[t]
\includegraphics[width=0.48\textwidth]{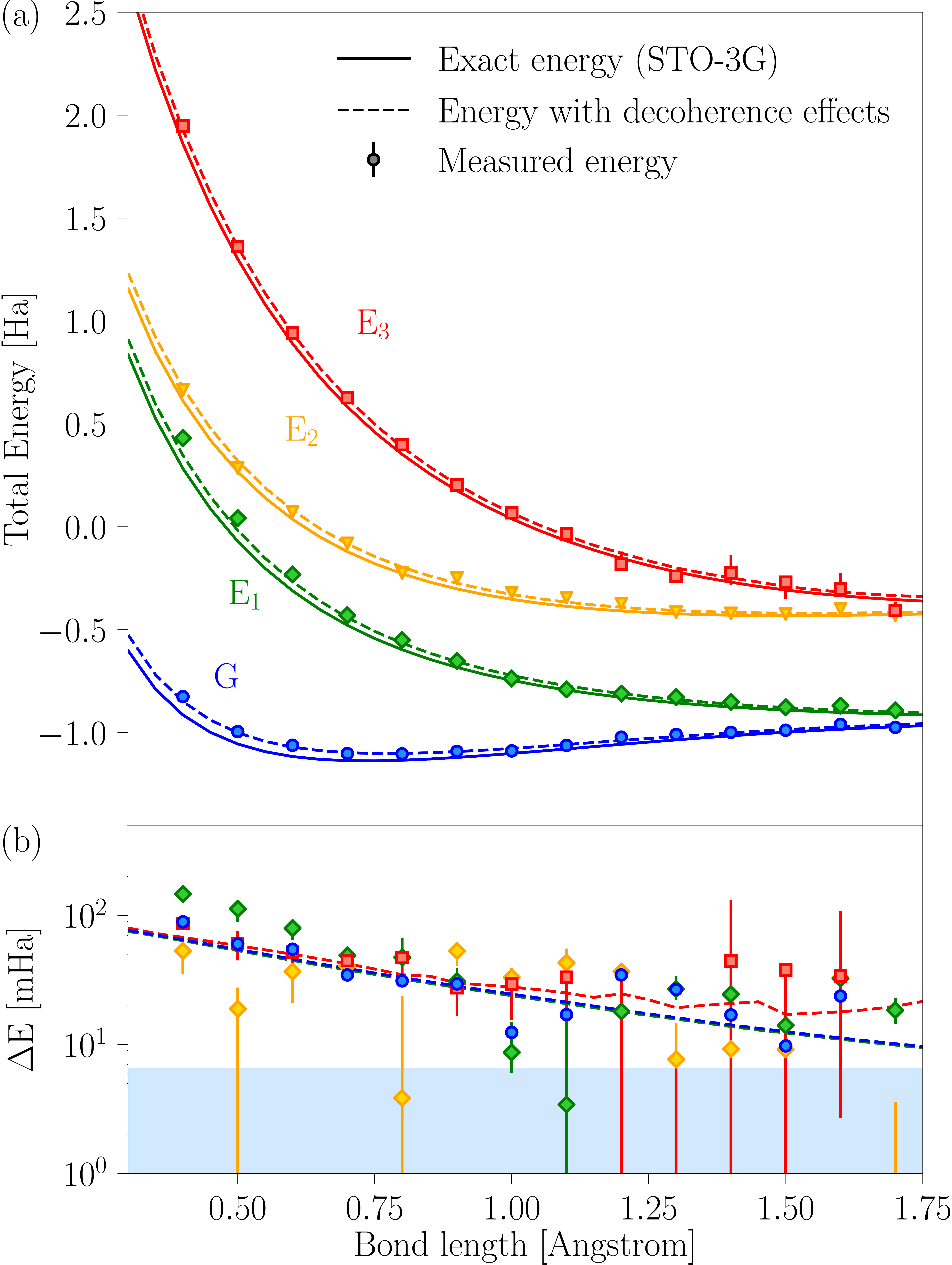}
\caption{Experimental VQE solution for the ground state and EOM solution for the excited states of molecular hydrogen using a tunable coupling architecture. (a) Ground (G) and excited state ($E_1$, $E_2$, $E_3$) energies as function of bond length. The symbols depict the experimental VQE solution, the solid lines represent the exact solution from the diagonalization of $\hat{H}_{\rm H_2}$, the dashed line represent the solution including decoherence effects. (b) Accuracy for ground and excited state energies as function of bond length. The symbols correspond to the accuracy of the measured ground and excited state energy determined with respect to the exact solution, while the dashed lines correspond to the expected accuracy including decoherence effects (see text). The depicted ground (excited) state energy is the minimum (median) value from a set of 5 measurements. The errorbars depict the range between the 1st and 3rd quantile (excited states only). The blue shaded area represents the region of chemical accuracy from 0 to 6.5 mHa. \label{fig:H2_curve}}
\end{figure}
\end{center}

To compute the ground state at a given bond length, the VQE searches for a state $|\psi(\vec{\theta})\rangle$ that minimizes the energy of the molecule $E(\vec{\theta})=\langle\psi(\vec{\theta})| \hat{H}_{\rm H_2} |\psi(\vec{\theta})\rangle$ using a classical optimization routine~\cite{Peruzzo2014}. First, an initial trial state $|\psi(\vec{\theta_0})\rangle$ is constructed and the energy $E(\vec{\theta_0})$ is calculated by measuring the expectation value $\langle\psi(\vec{\theta}_0)| \hat{O}_{i}\hat{O}_{j} |\psi(\vec{\theta}_0)\rangle$ of each term in Eq. \ref{eq:Hamiltonian} on the quantum hardware, where $\hat{O} = \{I,X,Y,Z\}$. Suitable trial states are of the form $|\psi(\vec{\theta})\rangle = a(\vec{\theta}) \ket{01} + b(\vec{\theta}) \ket{10}$ and can be realized in a single step with the exchange-type gate primitive $\hat{U}_\text{EX} (\theta,\varphi)$ (and a single initial qubit flip $\hat{X}_{\pi}$) in our tunable coupler architecture 
\begin{eqnarray}
\ket{\psi(\theta,\varphi)} &=& \hat{U}_{\text{EX}}(\theta,\varphi) \hat{X}_{\pi} \ket{00} \\
													 &=&	i e^{-i \varphi} \sin{(\theta/2)} \ket{01} + \cos{(\theta/2)} \ket{10}.
\end{eqnarray}
All trial states $\ket{\psi(\theta,\varphi)}$ are therefore probed from the one-excitation manifold by scanning the parameters $\theta(\tau)$ and $\varphi(\varphi_{\Phi})$. A simultaneous pertubation stochastic approximation (SPSA) algorithm is then used to determine a set of $(\theta_{\text{opt}},\varphi_{\text{opt}})$ corresponding to a state $\ket{\psi(\theta_{\text{opt}},\varphi_{\text{opt}})}$ that minimizes the energy $E(\theta_{\text{opt}},\varphi_{\text{opt}})$ for the given bond length (see Methods). By changing the parameters $\alpha_i$ in Eq. \ref{eq:Hamiltonian} and running the VQE again for the modified Hamiltonian, we compute the ground state energy of molecular hydrogen as a function of the bond length (Fig.~\ref{fig:H2_curve}). 

Furthermore, we compute the excited states of molecular hydrogen following the equation of motion (EOM) approach (see Methods). Using a variational method, we obtain a pseudo-eigenvalue system of equations which describes the excitations of the system. The matrix elements of this pseudo-eigenvalue system correspond to expectation values of a modified Hamiltonian with the ground state. For each bond length, we measure these matrix elements using the ground state $\ket{\psi(\theta_{\text{opt}},\varphi_{\text{opt}})}$ computed previously with VQE and solve the pseudo-eigenvalue system classically. The solution of this eigenvalue problem then yields the excited state energies.

For each bond length, we perform five runs of the experiment and plot the minimum value for the ground state energy and the median value for all excited state energies (symbols in Fig~\ref{fig:H2_curve}(a)). Comparing this experimental solution with the exact solution from a diagonalization of the Hamiltonian $\hat{H}_{\rm H_2}$ yields the accuracy $\Delta E$ (symbols in Fig.~\ref{fig:H2_curve}(b)).\\

\noindent
\textbf{Discussion}\\
For both ground and excited states, $\Delta E$ decreases with the bond length while staying above chemical accuracy (defined here by 6.5 mHa as in~\cite{Barkoutsos2018}). In order to understand this behavior, we study the influence of decoherence effects on the accuracy. Using the decoherence rates from Table \ref{tab:params} and a Lindblad-type master equation (see Methods), we simulate ground and excited state energies which now deviate from the exact solution due to decoherence effects (dashed lines in Fig.~\ref{fig:H2_curve}(a) and (b)). The numerical simulations are in good agreement with the experimental data indicating that decoherence has a strong influence on the measured accuracy in our experiment. In particular, the short coherence time $T_{\text{2,TC}}^{*} = 20~\rm{ns}$ of the tunable coupler in the present hardware is identified as the main cause of inaccuracy. Our simulations indicate that tunable couplers with coherence times of $T_{2,TC}^* = 500~\rm{ns}$ would enable us to reach chemical accuracy with the given architecture (see Supplementary material). We note that errors in the optimization and measurement of the ground state $\ket{\psi(\theta_{\text{opt}},\varphi_{\text{opt}})}$ can induce additional errors in the excited state energies. A detailed analysis of the different errors affecting the excited state calculation is beyond the scope of this work and will be discussed elsewhere~\cite{Ollitrault2018}. Also the question how the EOM compares with the QSE method~\cite{Colless2018,McClean2016} with respect to errors will be subject to future studies. 

In conclusion, we demonstrate a gate-efficient way to simulate molecular spectra on a tailor-made superconducting qubit processor using exchange-type two-qubit gates. With the choice of excitation-preserving exchange-type gates, tunable in both amplitude and phase, we preserve the number of excitations in the system and achieve the reduction of the VQE entangler to a single gate primitive. This enables the efficient computation of the molecular ground state, which can subsequently be used to efficiently calculate the molecule's excited states using an EOM approach. In the present case, the accuracy of the computation is still limited by the coherence time of the tunable coupling element. However, error mitigation schemes~\cite{Temme2016,Kandala2018} or minor improvements to the coherence of the coupler would allow us to reach chemical accuracy. Our findings show that adapting quantum algorithms and hardware to the problem at hand is a key requirement to perform quantum simulation on a larger scale. In particular, exchange-type gates are a promising choice to compute the energy spectra of larger molecules like water on near-term quantum hardware. \\%Since exchange-type operations naturally conserve the number of particles they are best suited to perform quantum chemistry calculation, where the number of electrons is constant. Here, we showed that the circuit depth of a VQE algorithm based on exchange-type gates is shorter compared with other two-qubit gates and calculate the energy spectrum of molecular hydrogen as a first application. Importantly, our approach indicates that it is essential to adapt both algorithms and hardware to efficiently solve the problem at hand. \textcolor{blue}{to be completed later}}\\%we have implemented a VQE algorithm based on exchange-type entangling gates to measure the energy spectrum of molecular hydrogen. Such exchange-type entanglers are best suited for quantum chemsitry calculation as they}%We highlight that exchange-type entanglers allow the design of quantum circuits with shorter circuit depth than other entanglers like the CNOT or CZ gate. This efficient generation of quantum circuits will allow the simulation of larger quantum systems like the water molecule on non-error corrected near-term quantum hardware. (@All: this not the final version, ill finish that part once the rest of the paper stands...)} \\

\section{Acknowledgments}
We thank the quantum team at IBM T. J. Watson Research Center, Yorktown Heights, in particular the fab team, Jerry Chow, David McKay and William Shanks for insightful discussions and the provision of qubit devices. We thank R. Heller and H. Steinauer for technical support. We thank M. Roth for theoretical support and discussions related to the tunable coupler architecture. This work was supported by the IARPA LogiQ program under contract W911NF-16-1-0114-FE and the ARO under contract W911NF-14-1-0124.\\

\section{Author contributions}
M.G: Design, implementation and analysis of the experiment, preparation of the manuscript. D.E: Significant contributions to design, implementation and analysis of the experiment. P.B: Development of the theory for gate-efficient quantum chemistry calculation. P.O. and S.W: Development of the equation of motion approach for the excited states calculation. G.S: Numerical simulation of accuracy including decoherence. N.M: Contributions to the theory of the tunable coupler and quantum chemistry calculation. A.F: Significant contributions to design and construction of the experimental setup. P.M: Design of dedicated electronics and contributions to the experimental setup. I.T: Significant contributions to all theoretical aspects of the work. S.F: Significant contributions to all theoretical and experimental aspects of the work. All authors contributed to the manuscript. \\

\noindent
\textbf{Methods}\\

%\noindent
%\textbf{Estimation of circuit depth and required hardware size}\\
%\textcolor{blue}{To be completed by Panos}

\noindent
\textbf{Classical optimization routine for VQE.}
The classical optimization of the VQE parameters ($\theta,\varphi$) is done by a simultaneous pertubation stochastic approximation (SPSA) algorithm~\cite{Spall1998} using the noisyopt python package. 

%As described in the main text, the parameters to be optimized are $\theta$ and $\varphi$. 
For the initial VQE parameters we use $\theta = \varphi = 0$ which corresponds to the initial state $|10\rangle$. Next, the SPSA parameters $\alpha$, $a$,$\gamma$ and $c$ need to be properly configured to ensure a fast convergence to the target state. Here, $\gamma$ and $c$ control the gradient approximation while $\alpha$ and $a$ control the update of the parameters $(\theta,\varphi)$ along that gradient. To ensure a robust gradient approximation we choose $\gamma = 0.101$ and $c=0.1$, such that $c$ is larger than the measured stochastic energy fluctuations $\epsilon \sim 0.02$ Ha. In previous VQE experiments~\cite{Kandala2017}, the parameter $\alpha=0.602$ was deemed optimal as it ensures a smooth convergence to the target. In the present experiment we choose a larger value $\alpha = 2$ to achieve a faster convergence to the target state. We finally calibrate the value $a$ as described in the supplementary information of~\cite{Kandala2017} and find an optimal value of $a = 20$. 

Convergence of the SPSA is achieved if the change in energy between consecutive iterations becomes smaller than the standard deviation in energy over the last 5 iterations. The optimized VQE parameters $(\theta_{\text{opt}},\varphi_{\text{opt}})$ are obtained after $16 \pm 4$ iterations with the SPSA configuration parameters described above. Using the optimized VQE parameters $(\theta_{\text{opt}},\varphi_{\text{opt}})$, a final measurement of all expectation values  $\langle\psi(\theta_{\text{opt}},\varphi_{\text{opt}})| \hat{O}_{i}\hat{O}_{j} |\psi(\theta_{\text{opt}},\varphi_{\text{opt}})\rangle$ with $\hat{O} = \{I,X,Y,Z\}$ and thus of the energy $E(\theta_{\text{opt}},\varphi_{\text{opt}})$ is performed.  \\

\noindent
\textbf{Numerical simulation of molecular energies including decoherence.}
To determine the influence of decoherence on the VQE accuracy, the time evolution of the system is calculated using a Lindblad-type master equation
\begin{eqnarray}
\dot{\rho} = -\frac{i}{\hbar} \left[\hat{H}_{tr},\rho \right] + \sum_{\text{i=Q1,Q2,TC}}{\Gamma_i^- \mathcal{L}[a_i] \rho+ \Gamma_i^z \mathcal{L}[a_i^{\dagger}a_i] \rho}
\label{Lindblad}
\end{eqnarray}
with the standard Lindblad operator $\mathcal{L}[C] = \left(2 \mathcal{C} \rho(t) \mathcal{C}^{\dagger} - \left\{ \rho(t),\mathcal{C}^{\dagger}\mathcal{C},\rho \right\}\right)/2 $. The decay rates for the i-th transmon are given by the dissipation rates reported in Table \ref{tab:params} via $\Gamma_i^z = (1/2)\left(1/T_{2,i}^{*} - 1/(2 T_{1,i}) \right)$ and $\Gamma_i^- = 1/T_{1,i}$. The master equation (Eq.~\ref{Lindblad}) with $\hat{H}_{tr}$ representing the coupled system of two transmons and a tunable coupler, implemented as three level systems, is numerically solved using QuTiP~\cite{Johansson2013a,Roth2017}. From this solution, the density matrix of the evolved state is calculated as a function of $\theta$ and $\phi$. For each bond length, energy values are obtained as the global minimum (with respect to $\theta$ and $\phi$) of the expectation value of the Hamiltonian $\hat{H}_{\rm H_2}$ defined in Eq.~\ref{eq:Hamiltonian}. The corresponding ground state $\ket{\psi(\theta_{\text{opt}},\varphi_{\text{opt}})}_{\text{sim}}$ is subsequently used to compute the excited state energies (see next Method section). \\ %From the obtained density matrix the expectation values $\langle\psi(\vec{\theta})| \sigma_{i}\sigma_{j} |\psi(\vec{\theta})\rangle_{\rm sim}$ can now be computed via ... \textcolor{blue}{@Gian: please complete and correct the above if not correct or incomplete}\\

\noindent
\textbf{Excited state calculation}\\
The calculation of the excited state is based on the minimization of the energy differences~\cite{Rowe1968}
\begin{equation}
\Delta E_{0n}= \frac{\langle \psi(\theta_{\text{opt}},\varphi_{\text{opt}})| [\hat O_n, \hat H, \hat O_n^{\dagger}] | \psi(\theta_{\text{opt}},\varphi_{\text{opt}}) \rangle}{\langle \psi(\theta_{\text{opt}},\varphi_{\text{opt}})| [\hat O_n, \hat O_n^{\dagger}]   | \psi(\theta_{\text{opt}},\varphi_{\text{opt}}) \rangle}
\label{Eq:deltaE}
\end{equation}
for a generic excitation/de-excitation operator $\hat O^\dagger_n=\sum_{\mu} \langle V_{\mu} \hat E_{\mu} - W_{\mu} \hat E^{\dagger}_{\mu} \rangle$ %with $\hat E_{\mu}=\hat a^{\dagger}_i \hat a_s$ and $\mu=(i,s)$ ($i$ labels occupied and $s$ unoccupied/virtual orbitals), and where $V_{\mu}$ and $W_{\mu}$ are the variational parameters. 
where $\hat E_{\mu}$ can be the fermionic creation/anihilation operators or the product of the two. The variable $\mu$ runs over $n$, the number of excited states to be computed. Here, $\hat{H}$ is the molecular Hamiltonian in second quantization and $[\hat A, \hat B, \hat C]=\frac{1}{2}\left([[\hat A,\hat B],\hat C]+[\hat A,[\hat B,\hat C]]  \right)$ where $[\hat A,\hat B]=\hat A\hat B-\hat B\hat A$.
In practice, it is sufficient to evaluate the ground state wavefunction $\ket{\psi(\theta_{\text{opt}},\varphi_{\text{opt}})}$ using for instance a VQE algorithm (see main text) and then evaluate the expectation values occurring in Eq.~\eqref{Eq:deltaE}. The variational problem $\delta \Delta E_{0n}=0$ in the variables $V_{\mu}$ and $W_{\mu}$ leads to a pseudo-eigenvalue system of equations of rank $2n$, the solutions of which are the excited state energies.

%\section{Bibliography}
%\bibliography{../IBMRefDB}

\end{document}

\title{Gate-efficient simulation of molecular eigenstates on a quantum computer}

\author{M. Ganzhorn, D.J. Egger, P. Barkoutsos, P. Ollitrault, G. Salis, N. Moll, A. Fuhrer, P. Mueller, S. Woerner, I. Tavernelli and S. Filipp}
\affiliation{IBM Research GmbH, Zurich Research Laboratory, S\"aumerstrasse 4, 8803 R\"uschlikon, Switzerland}

\date{\today}

\maketitle

\section{Estimation of circuit depth and required hardware size}

In this section we elaborate on the estimation of the circuit depth required to achieve chemical accuracy in the calculation of molecular eigenstates. To construct the circuits that perform the calculation, different methods can be used. Here we employ the heuristic approach with the use of excitation-conserving gates as described in~\cite{Barkoutsos2018} and the non-excitation-conserving hardware efficient approach introduced by~\cite{Kandala2017}. The convergence to the minimum energy values is achieved by using a Variational Quantum Eigensolver (VQE). 

\subsection{Heuristic excitation conserving circuits}
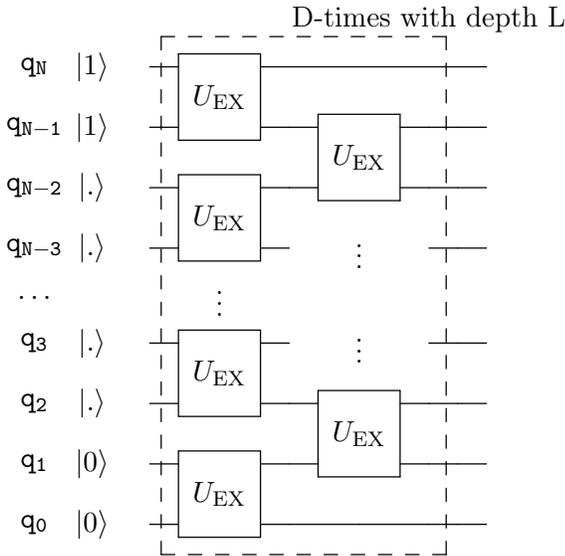
\begin{figure}[b!]
$$\begin{array}{c}
\Qcircuit @C=1em @R=1.2em {
&  & & & & & & & & &  \mbox{D-times with depth L}\\
\quad   & \quad & \mathtt{q_{N}}& \quad & \ket{1}  & \quad & \quad & \multigate{1}{U_{\rm{EX}}} & \qw &\qw  & \qw &\qw &\qw\\
\quad   & \quad & \mathtt{q_{N-1}}& \quad & \ket{1}  & \quad & \quad & \ghost{U_{\rm{EX}}}         & \qw & \multigate{1}{U_{\rm{EX}}} &\qw &\qw &\qw\\
\quad   & \quad & \mathtt{q_{N-2}}& \quad & \ket{.}  & \quad & \quad & \multigate{1}{U_{\rm{EX}}} & \qw & \ghost{U_{\rm{EX}}} &\qw &\qw &\qw\\
\quad   & \quad & \mathtt{q_{N-3}}& \quad & \ket{.}  & \quad & \quad & \ghost{U_{\rm{EX}}}         & \qw &  \quad \vdots \ \ \ & \quad &\qw &\qw\\
\quad   & \quad & \dots           & \quad  & \quad & \quad  & \quad        & \quad \vdots \ \ \  \\
\quad   & \quad & \mathtt{q_3}& \quad & \ket{.} & \quad & \quad & \multigate{1}{U_{\rm{EX}}} & \qw  &  \quad \vdots \ \ \ & \quad  &\qw &\qw\\
\quad   & \quad & \mathtt{q_2}& \quad & \ket{.}  & \quad & \quad & \ghost{U_{\rm{EX}}}         & \qw & \multigate{1}{U_{\rm{EX}}} &\qw &\qw &\qw\\
\quad   & \quad & \mathtt{q_1}& \quad & \ket{0}  & \quad & \quad & \multigate{1}{U_{\rm{EX}}} & \qw & \ghost{U_{\rm{EX}}} &\qw &\qw &\qw\\
\quad   & \quad & \mathtt{q_0}& \quad & \ket{0}  & \quad & \quad & \ghost{U_{\rm{EX}}}         & \qw  &\qw   & \qw &\qw &\qw
\gategroup{2}{8}{10}{11}{1.2em}{--}
}
\end{array}$$
\caption{Heuristic excitation-preserving circuit. \label{circuit:heuristic_SWAP}}
\end{figure}

On an $N$-qubit system, the quantum circuit typically consists of $D$ blocks of $N-1$ excitation-preserving gates $U_{\rm{EX}}$, with each individual block having an effective circuit depth of $L$ (Fig.~\ref{circuit:heuristic_SWAP}). Using QuTiP~\cite{Johansson2013a}, we estimate the minimum number of blocks $D_0$ required to achieve chemical accuracy for the ground state energy at the equilibrium point of each molecule, as described in Ref.~\cite{Barkoutsos2018}. We report the total circuit depth $D_0 \cdot L$ in Table~\ref{tab:Molecules_table_Exchange} and Fig. 1 of the main text. We assume error free gates and nearest neighbour connectivity, which is currently implemented in most hardware platforms~\cite{Kandala2017,Otterbach2017,omalley_scalable_2016}. All molecules are described in STO3G basis~\cite{Hehre1972} and a Jordan Wigner transformation is used for the mapping to qubit space. Additionally, effective core potentials~\cite{Barkoutsos2018} are used for LiH, BeH$_2$ and H$_2$O reducing the number of qubits by 2. We note that introducing gates errors and/or reducing the connectivity will result in a longer circuit depth. 

Furthermore, we calculate the effective runtime of the generated quantum circuit (Table~\ref{tab:Molecules_table_Exchange}), assuming gate times of 200 ns for CNOT gates~\cite{Sheldon2016b}, 170 ns for exchange type gates (see section for RB) and 25 ns for single qubit gates. 

\begin{table}[h!]
\caption{Statistics for the simulation of molecules and corresponding circuit depth and runtime.}
\label{tab:Molecules_table_Exchange}       % Give a unique label    
\centering
\begin{tabular}{ M{1.9cm} M{1.9cm}  M{1.9cm} M{2.5cm} M{2.5cm} M{2.5cm} M{2.5cm} }
 \hline
Molecule&  Qubits $N$ & Blocks $D_0$ & \multicolumn{2}{c}{Circuit Depth $D_0 \cdot L$} &\multicolumn{2}{c}{Circuit runtime  ($\mu s$)} \\ \hline
&   &   &Exchange & Decomp. &Exchange & Decomp. \\ \hline \hline
%H$_2$ (red.)   & 2    &1 & 1    & 9 &0.17 &1.125\\
H$_2$                   & 4    &4 &8 &72 &0.68 &4.5\\
LiH           		    & 10    & 14  &28 &252 &4.76 &31.5 \\
BeH$_2$  	    & 12  & 18 & 36 & 324 &6.12 &40.5 \\
H$_2$O		    & 12  & 20   &40 & 360 &6.8 &45 \\
 \hline
\noalign{\smallskip}\hline\end{tabular}

\end{table}
In order to compare with hardware architectures where exchange-type gates are not naturally available we also state the required circuit depth when decomposing an exchange-type gate into a sequence of two-qubit gate primities~\cite{Barenco1995} in Table~\ref{tab:Molecules_table_Exchange}. In the first step of the decomposition, a controlled change of basis ($R_z$) is performed followed by a general controlled rotation (in this case $R_x$) and then undoing the controlled of change basis. 

%{
%\begin{localsize}{12}{
$$\begin{array}{c}
\Qcircuit @C=1em @R=1.2em {
\quad   & \quad & \mathtt{q_j} & \quad  & \quad & \quad & \multigate{2}{U_{\rm{EX}}} & \qw  & \quad &\quad &\qw   &\ctrl{2}          &\gate{U_B(\phi)} &\ctrl{2} &\qw\\
& \quad & \quad & \quad        & \quad  & \quad        & \quad & \quad            & =    & \quad &\quad &\quad &\quad             &\quad              &\quad    &\quad\\
\quad   & \quad & \mathtt{q_i} & \quad  & \quad & \quad & \ghost{U_{\rm{EX}}}         & \qw  & \quad &\quad &\qw   &\gate{U_A(\theta)}&\ctrl{-2}          &\gate{U_C(\theta)}    &\qw\\
}
\end{array}$$
%}\end{localsize}}

and the $U_A$,$U_B$ and $U_C$ gates are 

%\begin{localsize}{12}{
\begin{equation}
\begin{array}{c c c}
\hat{U}_{\rm{A}}\left( \theta \right)=
  \begin{pmatrix}
    0          & e^{-i\frac{\theta}{2}}  \\
    e^{i\frac{\theta}{2}} & 0 \\
  \end{pmatrix},
&
\hat{U}_{\rm{B}}\left( \phi \right)=
  \begin{pmatrix}
    \cos \phi            & -i \sin \phi \\
    - i \sin \phi & \cos \phi\\
  \end{pmatrix},
&
\hat{U}_{\rm{C}}\left( \theta \right)=
  \begin{pmatrix}
    0          & e^{i\phi}  \\
    e^{i\phi} & 0 \\
  \end{pmatrix}
\end{array}
\end{equation}
%}\end{localsize}

To further decompose these gates in a CNOT primitive set of gates we need to follow the same procedure for every controlled rotation. The final circuit has the form 

%{
%\begin{localsize}{11}
$$\begin{array}{c}
\Qcircuit @C=0.4em @R=0.3em {
\quad   & \quad & \mathtt{q_j} & \quad  & \quad & \quad & \multigate{2}{U_{\rm{EX}}} & \qw  
& \quad & \qw & \qw & \ctrl{2} & \gate{R_z\left( \frac{-\pi}{2} \right)} & \gate{R_y \left( \frac{\phi}{2} \right)} &\targ & \gate{R_y \left( \frac{-\phi}{2} \right)} & \targ & \gate{R_z\left( \frac{\pi}{2} \right)}  & \ctrl{2} & \qw &\qw
\\
& \quad & \quad & \quad        & \quad  & \quad        & \quad & \quad & =    
& \quad\\
\quad   & \quad & \mathtt{q_i} & \quad  & \quad & \quad & \ghost{U_{\rm{ex}}} & \qw 
& \quad & \qw & \gate{R_z \left( \frac{\theta}{2} \right)} & \targ & \gate{R_z\left( \frac{-\theta}{2} \right)} &\qw
&\ctrl{-2} &\qw & \ctrl{-2} & \gate{R_z \left( \frac{-\theta}{2} \right)} & \targ & \gate{R_z\left( \frac{\theta}{2} \right)} &\qw \\
}
\end{array}$$
%\end{localsize}
%}

Using this gate transformation we can substitute the $U_{\rm EX}$ with a series of single and two qubit rotations that result to the same circuit, but with an ninefold increase in the circuit depth. Nevertheless, for the studied molecules the circuit runtime is still within the $T_1$ limit (see Fig. 1, main text) which would allow an implementation on the corresponding quantum hardware. 

\subsection{Heuristic non excitation-conserving circuits}

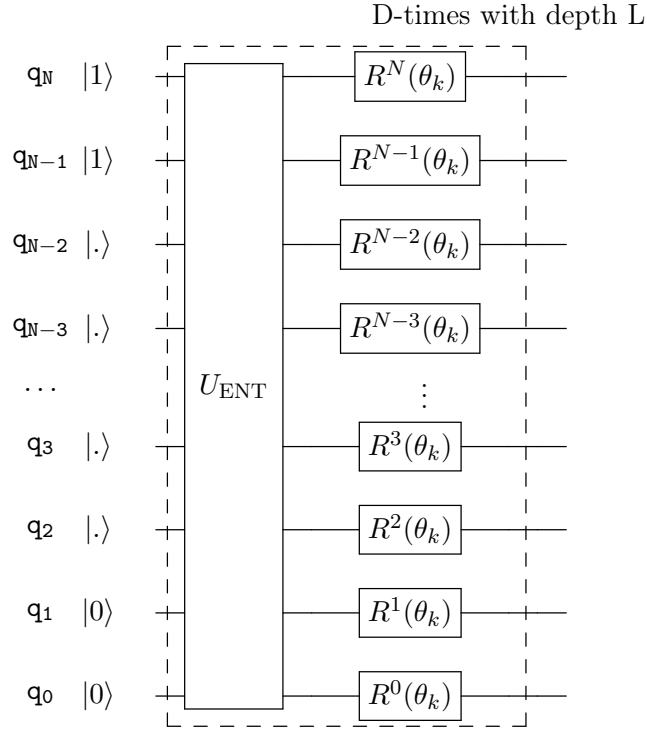
\begin{figure}[b!]
$$\begin{array}{c}
\Qcircuit @C=1em @R=1.2em {
&  & & & & & & & & &  \mbox{D-times with depth L}\\
\quad   & \quad & \mathtt{q_{N}}	& \quad & \ket{1}  & \quad & \quad & \multigate{8}{U_{\rm{ENT}}}& \qw 	& \gate{R^N(\theta_k)} 			& \qw &\qw &\qw\\
\quad   & \quad & \mathtt{q_{N-1}}& \quad & \ket{1}  & \quad & \quad & \ghost{U_{\rm{ENT}}} 			& \qw 	& \gate{R^{N-1}(\theta_k)}  & \qw &\qw &\qw\\
\quad   & \quad & \mathtt{q_{N-2}}& \quad & \ket{.}  & \quad & \quad & \ghost{U_{\rm{ENT}}} 			& \qw 	& \gate{R^{N-2}(\theta_k)}  & \qw &\qw &\qw\\
\quad   & \quad & \mathtt{q_{N-3}}& \quad & \ket{.}  & \quad & \quad & \ghost{U_{\rm{ENT}}} 			& \qw 	& \gate{R^{N-3}(\theta_k)}  & \qw &\qw &\qw\\
\quad   & \quad & \dots           & \quad & \quad 	 & \quad & \quad & \quad  										& \quad	& \quad	\vdots							&\quad&\quad &\quad\\
\quad   & \quad & \mathtt{q_3}		& \quad & \ket{.}  & \quad & \quad & \ghost{U_{\rm{ENT}}} 			& \qw 	& \gate{R^3(\theta_k)}  		& \qw &\qw &\qw\\
\quad   & \quad & \mathtt{q_2}		& \quad & \ket{.}  & \quad & \quad & \ghost{U_{\rm{ENT}}} 			& \qw 	& \gate{R^2(\theta_k)}  		& \qw &\qw &\qw\\
\quad   & \quad & \mathtt{q_1}		& \quad & \ket{0}  & \quad & \quad & \ghost{U_{\rm{ENT}}} 			& \qw 	& \gate{R^1(\theta_k)}  		& \qw &\qw &\qw\\
\quad   & \quad & \mathtt{q_0}		& \quad & \ket{0}  & \quad & \quad & \ghost{U_{\rm{ENT}}} 			& \qw 	& \gate{R^0(\theta_k)}  		& \qw &\qw &\qw
\gategroup{2}{8}{10}{11}{1.2em}{--}
}
\end{array}$$
\caption{Heuristic non-excitation-preserving circuit. \label{circuit:heuristic_CNOT}}
\end{figure}

To simulate molecules with non excitation-conserving circuits we implement the scheme introduced by Kandala et. al. in Ref.~\cite{Kandala2017} using CNOT gates primitives. In this heuristic approach, the quantum circuits are constituted of $D$ blocks containing single qubit rotations $R^n(\theta_k)$ and an entangler circuit $U_{\rm{ENT}}$ based on CNOT operations, with each block having an effective circuit depth of $L$ (Fig.~\ref{circuit:heuristic_CNOT}). Using the open source library QISKIT AQUA Chemistry~\cite{qiskit_aqua_url}, we estimate the minimum number of blocks $D_0$ required to reach chemical accuracy for the ground state energy at the equilibrium point of each molecule and report the total circuit depth $D_0 \cdot L$ and effective circuit runtime in Table \ref{tab:Molecules_table_CNOT}. The simulation assumes error free gates and all-to-all connectivity, making these requirements the best case scenario for the corresponding computations. Errors in the gates and smaller connectivity will result in larger circuit depth. 

\begin{table}[h!]
\caption{Statistics for the simulation of molecules and corresponding circuit depth and runtime.}
\label{tab:Molecules_table_CNOT}       % Give a unique label    
\centering
\begin{tabular}{ M{1.9cm} M{1.6cm}  M{2cm}M{2cm} M{2cm} M{2cm}}
 \hline
Molecule&  Qubits $N$ &  Blocks $D_0$ & Circuit Depth $D_0 \cdot L$ & Circuit runtime ($\mu s$) \\ \hline\hline
%&   &   HF state & $\ket{0}^{\otimes N}$ state & HF state & $\ket{0}^{\otimes N}$ state &HF state & $\ket{0}^{\otimes N}$ state \\ \hline \hline
%H$_2$ (red.)         & 2    & 1 & 1 & 3    & 3 &0.425   &0.425\\
H$_2$                   & 4    & 5    & 80  & 7.3 \\ %& 4    & 5   &  6   & 80  & 96  & 7.3  & 8.7 \\
LiH           		    & 10  & 14 & 1890  & 208.9 \\%& 10  & 14 & 16  & 1890  & 2160  & 208.9 & 238.8\\
BeH$_2$  	    & 12  &>26  & >5850 & >626\\%& 12  & >26 &>26  & >5850 & >5850 & >626 & >626\\
H$_2$O		    & 12  & >26 & >5850 & >626\\%& 12  & >26 &>26  & >5850 & >5850 & >626 & >626\\
 \hline
\noalign{\smallskip}\hline\end{tabular}

\end{table}

The calculations for LiH molecule indicate that for acquiring the ground state properties within chemical accuracy the circuit runtime exceeds the available $T_1$ times (see Fig. 1, main text). 
These results indicate that the non-excitation conserving heuristic approach, even for error free gate simulations, yield circuits that are too long, making the simulation intractable in near-term hardware.
For BeH$_2$ and H$_2$O molecules we did not find a solution within chemical accuracy for circuits with depth up to 26 blocks within a reasonable computation time.
Further increase in the number of blocks would probably allow for convergence, but the obtained circuit depths will be prohibitive in the available hardware. 

\section{Coherence measurements and flux noise characterization}
In this section, we report on the measured relaxation and coherence times $T_1$, $T_2$, $T_2^*$ of the device used in these experiments
\subsection{Tunable coupler transmon}
The tunable coupler (TC) is implemented as a flux-tunable transmon (see main text). Consequently, its coherence time is affected by magnetic flux noise with a power spectral density of the form $S(\omega) = A^2/\omega$. According to Ref.~\cite{McKay2016}, the relationship between the TC dephasing time $T_{\phi}$ and flux noise is given by
\begin{equation}
T_{\phi} = \frac{2 T_1 T_2^*}{2 T_1 - T_2^*} = \frac{1}{A} \left|\frac{\partial \omega_{\text{c}}}{\partial \Phi} \right|^{-1} + T_{\text{other}}
\label{eq:Tphi}
\end{equation}
where $\omega_{\text{c}} = \omega_\text{c}^0\sqrt{\abs{\cos(\pi\Phi/\Phi_0)}}$ is the TC frequency and $T_{\text{other}}$ accounts for all other noise sources. Measuring $T_1$ and $T_2^*$ of the tunable coupler as function of the magnetic flux thus provides an estimate flux noise magnitude $A$. 
\begin{center}
\begin{figure}[hbt!]
\includegraphics[width=0.8\textwidth]{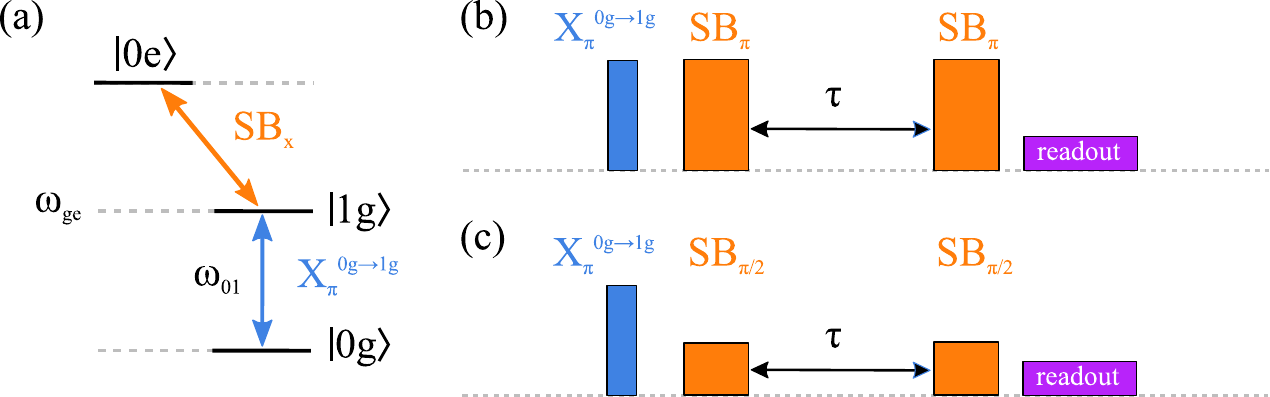}
\caption{Measuring the coherence of the tunable coupler. (a) Energy level scheme showing the population transfer between qubit and TC via a sideband (SB) drive. The parameter $x$ denotes the amount of population transfered in the process. (b) Pulse sequence for measuring the relaxation time $T_1$ of the TC. (c) Pulse sequence for measuring the relaxation time $T_2^*$ of the TC. \label{fig:coherence_meas_tc}}
\end{figure}
\end{center}
Due to the absence of a dedicated readout resonator for the TC, the population in the TC can only be measured indirectly via one of the computational qubits. Furthermore, a direct excitation of the TC via the fluxline is not possible due to heavy filtering at the frequency $\omega_{\text{c}} = \omega_{ge}$. However, parametric driving of the TC at the frequency $\omega = \omega_{ge} - \omega_{01}$ allows a population $x$ to be transfered from the state $|1g\rangle $ to a superposition state $\cos(x/2)|1g\rangle + \sin(x/2)|0e\rangle$ (orange transition in Fig.~\ref{fig:coherence_meas_tc}(a)). Here, $|0\rangle$ ($|1\rangle$) denotes the ground (excited) state of the qubit, while $|g\rangle$ ($|e\rangle$) denotes the ground (excited) state of the TC. In particular for $x=\pi$ a full population transfer $\rm SB_{\pi}$ from $|1g\rangle $ to $|0e\rangle$ can be achieved. Similarly, setting $x=\pi/2$ ($\rm SB_{\pi/2}$) creates a superposition $1/\sqrt{2}|1g\rangle + 1/\sqrt{2}|0e\rangle$.
%However, by driving the transition $|1g\rangle \rightarrow |0e\rangle$ population can be transfered between the qubit and the TC. This can be achieved by parametric driving of the TC at a frequency $\omega = \omega_{10} - \omega_{ge}$. 
With the available operations described above, the $T_1$ and $T_2^*$ times of the TC can be measured using the pulse sequences in Fig.~\ref{fig:coherence_meas_tc}(b) and (c), respectively. 
\begin{center}
\begin{figure}[hbt!]
\includegraphics[width=0.99\textwidth]{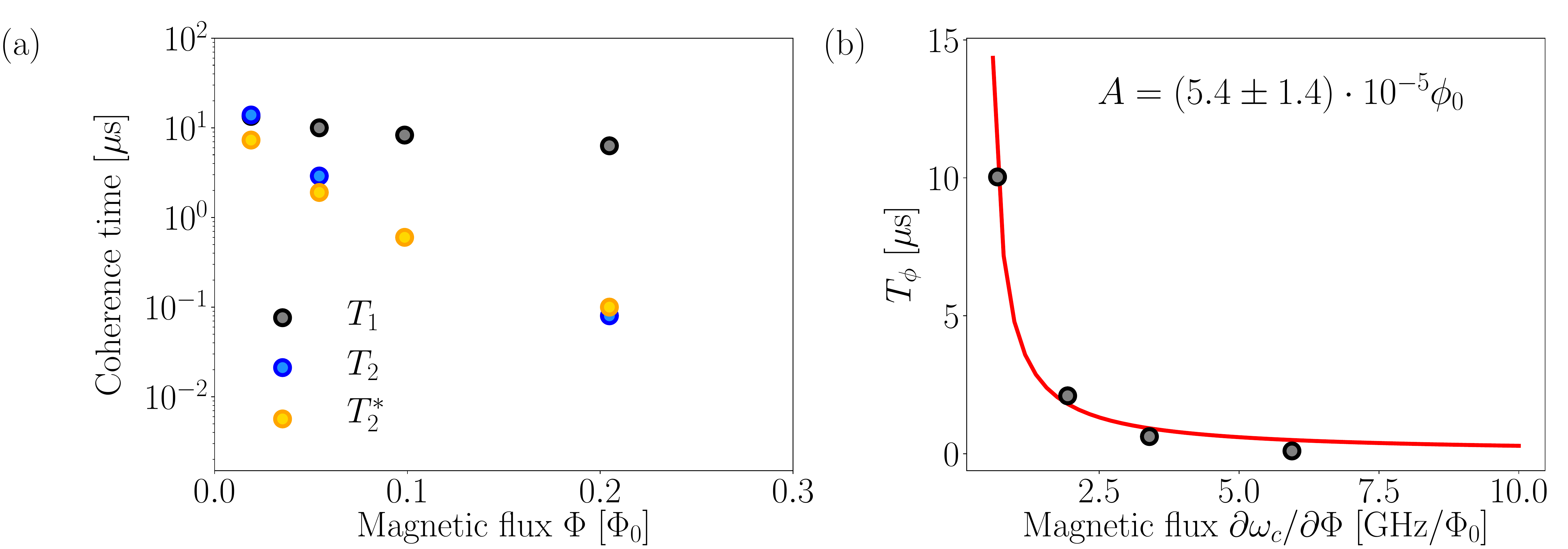}
\caption{Coherence of the tunable coupler. (a) Measurement of $T_1$, $T_2$ and $T_2^*$ time as function of magnetic flux. (b) $T_{\phi}$ as function of $\partial \omega_c / \partial \Phi$. The red solid line is a fit with equation \ref{eq:Tphi}. \label{fig:coherence_tc}}
\end{figure}
\end{center}
Based on the measured coherence times (see Fig.~\ref{fig:coherence_tc}), we find a flux noise amplitude of $A = 5.4 \pm 1.4 \cdot 10^{-5}$ $\Phi_0$ with a SQUID loop size of the tunable coupler of $S= 25 \cross 25$ $\mu$m$^2$%, 
\subsection{Computational transmon qubits}
\begin{center}
\begin{figure}[hbt!]
\includegraphics[width=0.99\textwidth]{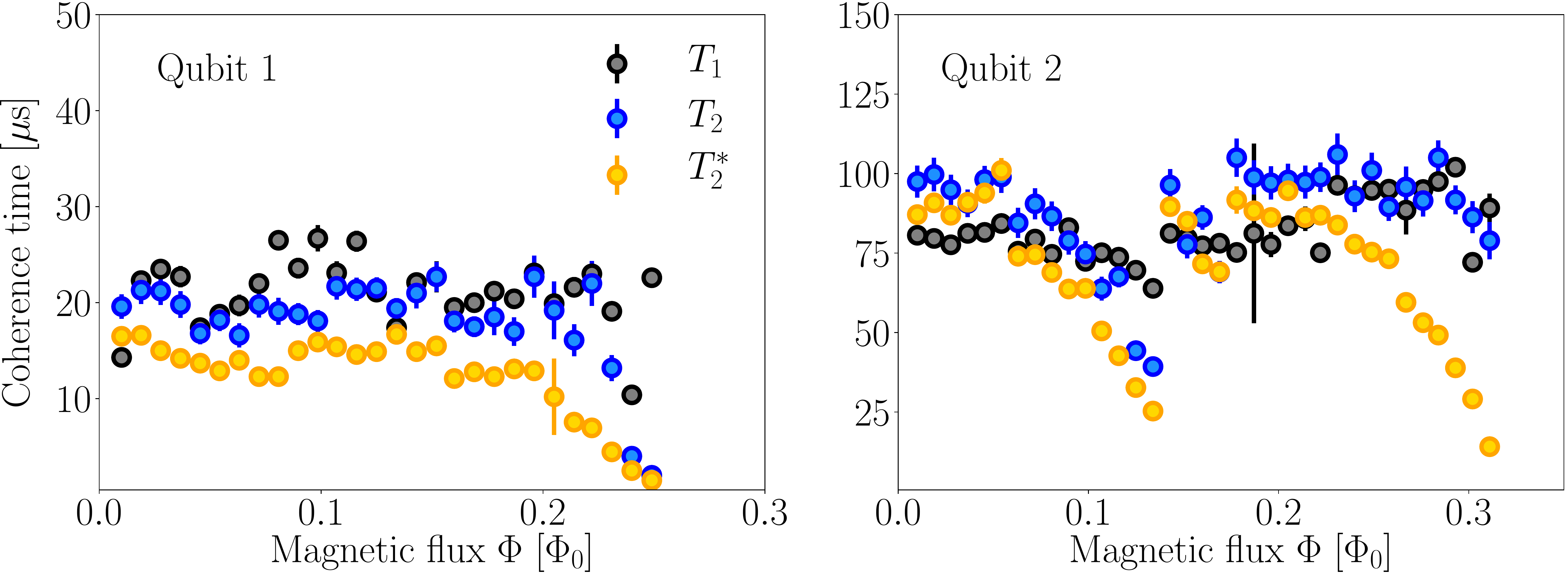}
\caption{Measurement of coherence time vs. magnetic flux for both computational qubits. \label{fig:coherence}}
\end{figure}
\end{center}

In the tunable coupler architecture, the computational qubits are capacitively coupled to the tunable transmon and are thus also subject to flux noise. $T_1$, $T_2$ and $T_2^*$ times of both computational qubits Q1 and Q2 as a function of magnetic flux $\Phi$ are shown in Fig. \ref{fig:coherence}. 

\section{Randomized benchmarking of iSWAP gates in TC architecture}
\begin{center}
\begin{figure}[hbt!]
\includegraphics[width=0.48\textwidth]{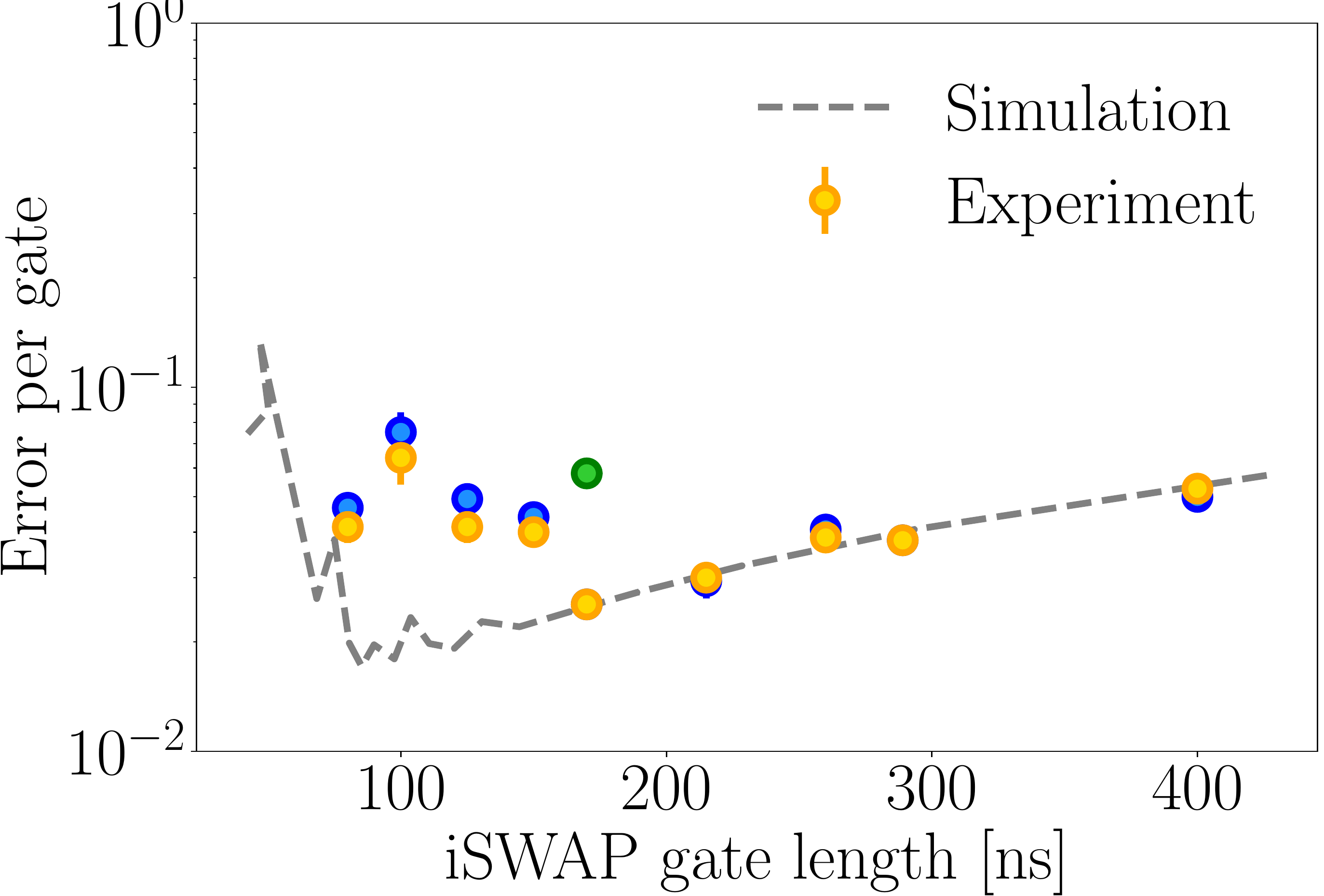}
\caption{Randomized benchmarking of the iSWAP gate. Blue (orange) dots depict RB measurement probing the ground state of qubit 1 (qubit 2). Experimental data are compared with numerical simulations (grey dashed lines) and error estimates via QPT measurements (green dot). \label{fig:RB}}
\end{figure}
\end{center}

While $\hat{U}_\text{EX}$ is in general not part of group of Clifford gates, the iSWAP gate defined by $\hat{U}_\text{EX}(\theta=\pi, \varphi=0)$ is a Clifford gate. Therefore it can be characterized via randomized benchmarking (RB). Fig.~\ref{fig:RB} shows the error per gate (EPG) as a function of the length of the iSWAP gate. Note that there are two sets of experimental data; one is obtained by measuring the ground state of qubit 1 (blue dots) and the other one by measuring the ground state of qubit 2 (orange dots).  

Furthermore, we estimate the iSWAP EPG using a Lindblad type master equation and the decoherence rates from Table 1 in the main text as described in ref.~\cite{Roth2017} (grey dashed line in Fig.~\ref{fig:RB}). For longer iSWAP gates the experimental findings agree well with the simulation, indicating that the EPG is limited by the coherence time of the tunable coupler $T_{\text{2,TC}}^{*} = 20$ ns. At short gate length, we observe an increase in the EPG due to leackage out of the computational subspace~\cite{Roth2017}. 
%Finally, we compare the EPG from RB with the EPG obtained via QPT in the main text for a gate length of 170 ns (green dot). Due to SPAM errors, the EPG from QPT is larger than the EPG resulting from RB.

\section{Numerical simulations of the VQE accuracy including decoherence effects}
\begin{center}
\begin{figure}[hbt!]
\includegraphics[width=0.48\textwidth]{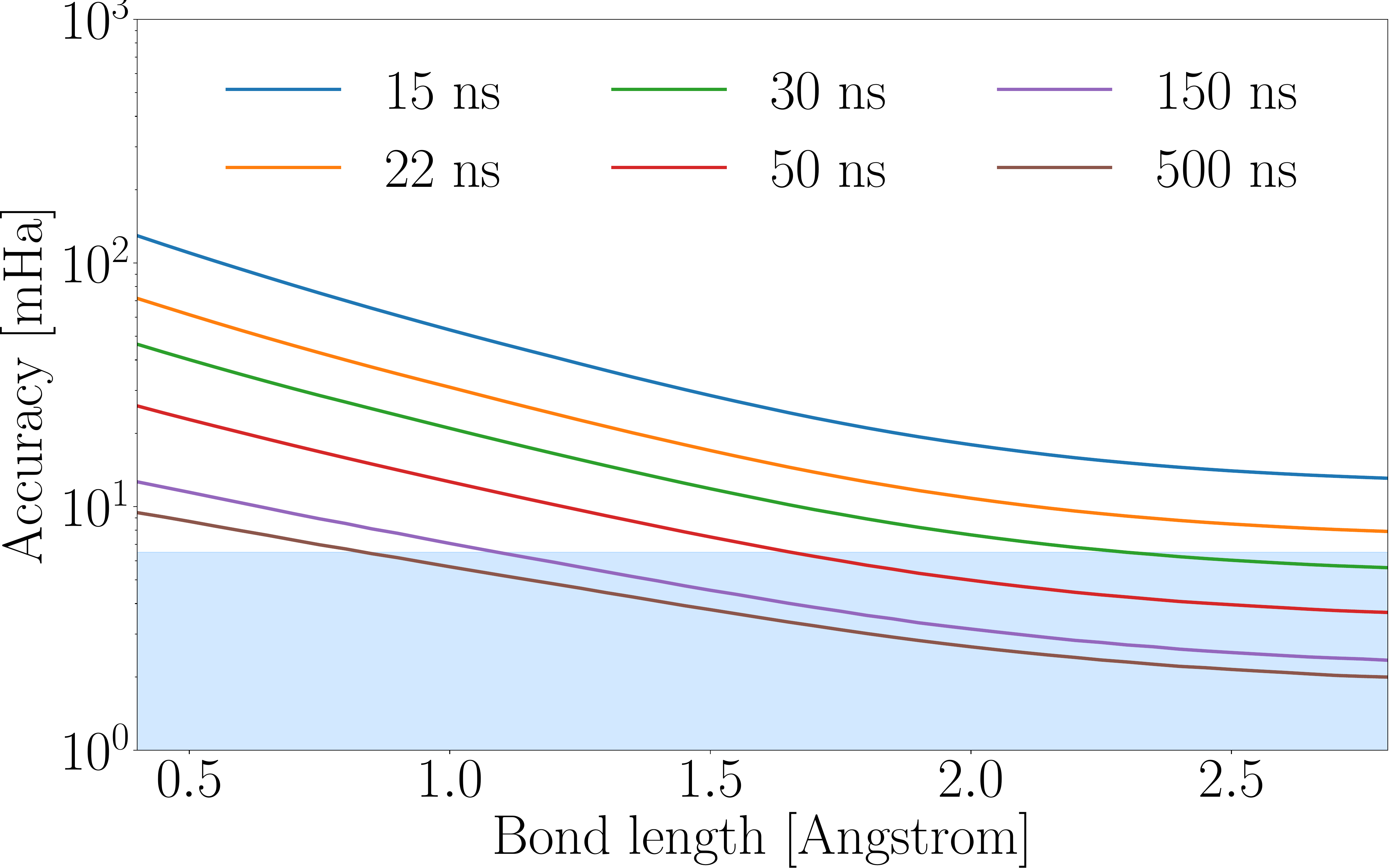}
\caption{Accuracy of the ground state VQE solution for molecular hydrogen as function of the bond length for different $T_2^*$ of the TC. \label{fig:TCaccuracy}}
\end{figure}
\end{center}
In this section, we show the influence of the $T_2^*$ time of the tunable coupler on the accuracy of the VQE solution. 

For a given bond length, we solve a Lindblad-type master equation as described in the Methods section of main text. From this solution, we obtain the density matrix of the evolved state $\ket{\psi (\theta,\varphi)}$ as a function of $\theta$ and $\varphi$ and evaluate the corresponding expectation values $\langle\psi(\theta,\varphi)| \hat{H}_{H_2} |\psi(\theta,\varphi)\rangle$. A global minimum search in the space spaned by all $(\theta,\varphi)$ yields the ground state energy for the given bond length. From the latter, we obtain the VQE accuracy by subtracting the energy obtained from a diagonalization of the Hamiltonian as described in the main text.

In Fig.~\ref{fig:TCaccuracy}, we compare the accuracy of the VQE for different $T_2^*$ times of the tunable coupler. The simulations indicate that chemical accuracy can be achieved for $T_2^* > 500$ ns.

\section{Reduced Hamiltonian for molecular hydrogen}
\begin{table}[hbt!]
\centering
\resizebox{0.48\textwidth}{!}{%
\begin{tabular}{c|ccccc}
\hline\hline
R(\AA)   	 & $\alpha_0$ 								 & $\alpha_1$    								& $\alpha_2$ 										& $\alpha_3$ 						 		& $\alpha_4$ \\\hline\hline
0.30 &-0.75374 	&  0.80864	& -0.80864	& -0.01328	& 0.16081 \\
%0.35 &-8.10661826e-01&  7.47415789e-01& -7.47415789e-01& -1.31035849e-02& 1.62573223e-01 \\
0.40 &-0.86257	&  0.68881	& -0.68881	& -0.01291	& 0.16451 \\
%0.45 &-9.08402086e-01&  6.33889808e-01& -6.33889808e-01& -1.27191769e-02& 1.66621403e-01 \\
0.50 &-0.94770	&  0.58307	& -0.58307	& -0.01251	& 0.16887 \\
%0.55 &-9.80513852e-01&  5.36488826e-01& -5.36488826e-01& -1.23003469e-02& 1.71244516e-01 \\
0.60 &-1.00712	&  0.49401	& -0.49401	& -0.01206	& 0.17373 \\
%0.65 &-1.02805039e+00&  4.55433445e-01& -4.55433445e-01& -1.18019177e-02& 1.76318450e-01 \\
0.70 &-1.04391	&  0.42045	& -0.42045	& -0.01150	& 0.179005 \\
%0.75 &-1.05540303e+00&  3.88747614e-01& -3.88747614e-01& -1.11771448e-02& 1.81771535e-01 \\
0.80 &-1.06321	&  0.35995	& -0.35995	& -0.01080	& 0.18462 \\
%0.85 &-1.06798505e+00&  3.33746523e-01& -3.33746523e-01& -1.04060675e-02& 1.87561845e-01 \\
0.90 &-1.07028	&  0.30978	& -0.30978	& -0.00996	& 0.19057 \\
%0.95 &-1.07057705e+00&  2.87796026e-01& -2.87796026e-01& -9.50347220e-03& 1.93650309e-01 \\
1.00 &-1.06924	&  0.26752	& -0.26752	& -0.00901	& 0.19679 \\
%1.05 &-1.06657841e+00&  2.48783316e-01& -2.48783316e-01& -8.50993897e-03& 1.99984260e-01 \\
1.10 &-1.06281	&  0.23139	& -0.23139	& -0.00799	& 0.20322 \\
%1.15 &-1.05812757e+00&  2.15233961e-01& -2.15233961e-01& -7.47720241e-03& 2.06494669e-01 \\
1.20 &-1.05267	&  0.20018	& -0.20018	& -0.00696	& 0.20979 \\
%1.25 &-1.04656498e+00&  1.86173128e-01& -1.86173128e-01& -6.45559441e-03& 2.13102394e-01 \\
1.30 &-1.03991 	&  0.17310	& -0.17310	& -0.00596	& 0.21641 \\
%1.35 &-1.03281974e+00&  1.60926414e-01& -1.60926414e-01& -5.48621795e-03& 2.19727027e-01 \\
1.40 &-1.02535	&  0.14956	& -0.14956	& -0.00503	& 0.22302 \\
%1.45 &-1.01761101e+00&  1.38976800e-01& -1.38976800e-01& -4.59758636e-03& 2.26294251e-01 \\
1.50 &-1.00964	&  0.12910	& -0.12910	& -0.00418	& 0.22953 \\
%1.55 &-1.00152212e+00&  1.19893558e-01& -1.19893558e-01& -3.80557852e-03& 2.32740283e-01 \\
1.60 &-0.99329	&  0.11130	& -0.11130	& -0.00344	& 0.23590 \\
%1.65 &-9.85022658e-01&  1.03305348e-01& -1.03305348e-01& -3.11546011e-03& 2.39013637e-01 \\
1.70 &-0.97673	&  0.09584	& -0.09584	& -0.00280	& 0.24207 \\
%1.75 &-9.68479302e-01&  8.88905676e-02& -8.88905676e-02& -2.52481570e-03& 2.45075012e-01 \\
1.80 &-0.96028	&  0.08240	& -0.08240	& -0.00226	& 0.24801 \\
\hline\hline
\end{tabular}}
\caption{Prefactors defining the Hamiltonian $H_{\rm H_2}$ in the STO-3G basis for different bond length $R$ \label{tab:params}}
\end{table}
For the calculation of the states of molecular hydrogen on a two-qubit device, we use the fermionic Hamiltonian in second quantization
\begin{eqnarray}
\hat{H} =\sum_{\alpha, \beta=0}^{M} t_{\alpha \beta} \, \hat{a}^\dagger_{\alpha} \hat{a}_{\beta} +\frac{1}{2}  \sum_{\alpha, \beta, \gamma, \delta = 1}^{M} u_{\alpha \beta \gamma \delta}\, \hat{a}^\dagger_{\alpha} \hat{a}^\dagger_{\gamma} \hat{a}_{\delta} \hat{a}_{\beta},
\end{eqnarray}
with the fermionic creation (annihilation) operator $a^\dagger_\alpha$ $(a_\alpha)$ for the molecular orbital $\alpha$. For the $H_2$ molecule, the number of molecular orbitals is $M=4$ and the one- and two-body interactions terms yield:
\begin{eqnarray}
t_{\alpha\beta}=\int d\boldsymbol x_1\Psi_\alpha(\boldsymbol{x}_1) \left(-\frac{\boldsymbol\nabla_1^2}{2}+\sum_{i} \frac{Z_i}{|\boldsymbol{r}_{1i}|}\right)\Psi_\beta (\boldsymbol{x}_1),\\
u_{\alpha\beta\gamma\delta}=\int\int d \boldsymbol{x}_1 d \boldsymbol{x}_2 \Psi_\alpha^*(\boldsymbol{x}_1)\Psi_\beta(\boldsymbol{x}_1)\frac{1}{|\boldsymbol{r}_{12}|}\Psi_\gamma^*(\boldsymbol{x}_2)\Psi_\delta(\boldsymbol{x}_2)
\end{eqnarray}
where we have defined the nuclei charges $Z_i$, the nuclei-electron and electron-electron separations $\boldsymbol{r}_{1i}$ and $\boldsymbol{r}_{12}$, the $\alpha$-th orbital wavefunction $\Psi_\alpha(\boldsymbol{x}_1)$, and we have assumed that the spin is conserved in the spin-orbital indices $\alpha,\beta$ and $\alpha,\beta,\gamma,\delta$.

We then map the fermionic Hamiltonian $\hat{H}$ to the qubit Hamiltonian $\hat{H}_{\rm H_2}$ using a parity mapping transformation~\cite{Bravyi2017}
\begin{eqnarray}
\hat{H}_{\rm H_2} = \alpha_0 II + \alpha_1 ZI + \alpha_2 IZ + \alpha_3 ZZ + \alpha_4 XX
\label{eq:Hamiltonian}
\end{eqnarray}
Here, the pre-factors $\alpha_i$ are a function of the one- and two-body interaction terms $t_{\alpha\beta}$ and $u_{\alpha\beta\gamma\delta}$, which can be computed by the software PyQuante~\cite{PyQuante}. The numerical values of the pre-factors $\alpha_i$ are summarized in Table~\ref{tab:params} for different bond length of the molecule.